\documentclass{article}

\usepackage{PRIMEarxiv}

\usepackage[utf8]{inputenc} % allow utf-8 input
\usepackage[T1]{fontenc}    % use 8-bit T1 fonts
\usepackage{hyperref}       % hyperlinks
\usepackage{url}            % simple URL typesetting
\usepackage{booktabs}       % professional-quality tables
\usepackage{amsfonts}       % blackboard math symbols
\usepackage{nicefrac}       % compact symbols for 1/2, etc.
\usepackage{microtype}      % microtypography
\usepackage{lipsum}
\usepackage{fancyhdr}       % header
\usepackage{graphicx}       % graphics
\graphicspath{{media/}}     % organize your images and other figures under media/ folder

\usepackage{amsmath}
\usepackage{stmaryrd}
\usepackage{tikz}
\usepackage{amssymb}
\usetikzlibrary{shapes.geometric, arrows, positioning, calc}
\usepackage{amsthm}

\title{Separation Logic for Memory Conflict Detection in High-Level Synthesis
}

\author{
  Yeonseok Lee \\
  SLING AI Inc. \\
  Incheon, Republic of Korea \\
  \texttt{ylee@sling.ai.kr}
}

\begin{document}

\newcommand{\bigast}{\scalebox{1.5}{$\ast$}}
\newcommand{\bank}{\textsf{bank}}
\newcommand{\arr}{\textsf{arr}}
\newcommand{\cfu}{\textsf{cfu}}

\maketitle

\begin{abstract}
High-Level Synthesis leverages loop unrolling and array partitioning, but scheduling concurrent accesses is challenging when indices contain non-affine arithmetic. Conventional polyhedral frameworks systematically over-approximate these non-linear transformations, forcing conservative serialization that degrades performance. To minimize this bottleneck, we present a spatial verification framework operating at the LLVM Intermediate Representation (IR) level. By extracting flat arithmetic expressions from ``getelementptr'' instructions, it models memory banks as polymorphic spatial predicates to handle non-affine terms. Structural safety is enforced via a Conflict-Free Unrolling condition using Separation Logic's separating conjunction; concurrent operations targeting the same bank trigger an automatic spatial contradiction. This disjointness requirement is reduced to a matrix of pairwise inequalities over immutable Static Single Assignment (SSA) variables for a Satisfiability Modulo Theories (SMT) oracle. To guarantee safety against undecidable non-linear arithmetic, we implement a deterministic sequential fallback. Finally, a theorem of soundness bridges algebraic SMT verification with Register Transfer Level trace safety, ensuring physical hardware immune to structural memory collisions.
\end{abstract}


\section{Introduction}

\subsection{The Drive for Hardware Parallelism in HLS}
High-Level Synthesis (HLS) drastically shortens the hardware development cycle by shifting design entry from structural Register Transfer Level (RTL) syntax up to high-level programming languages like C/C++ \cite{cong2011high,nane2015survey}. 
Modern state-of-the-art HLS frameworks leverage compiler optimization pipelines to extract loop-level parallelism and automatically schedule operations into efficient parallel hardware architectures \cite{canis2013legup,josipovic2018dynamically}. 
To fully match this computational parallelism, the underlying memory system must deliver high bandwidth, which FPGAs achieve by partitioning and distributing application data across multiple isolated on-chip memory blocks or banks \cite{wang2013memory}. 
Loop unrolling duplicates sequential basic blocks to expose concurrent iteration traces within a single clock cycle, but its throughput gains are fundamentally bounded by the memory infrastructure's ability to support non-interfering parallel data accesses \cite{winterstein2015separation}.

\subsection{The Bottleneck: Memory Conflicts and Affine Limitations}
Concurrently scheduling these duplicated memory instructions incurs a significant risk of structural hazards when multiple parallel operations attempt to access the same single-ported memory bank during the same clock cycle. To track array coordinates within multi-dimensional iteration spaces, standard dependency solvers historically employ the polyhedral model \cite{bondhugula2008practical}. However, these geometric techniques are bounded to static control structures where loop bounds and array access functions are affine combinations of enclosing loop variables.

In the presence of pointer arithmetic, dynamic memory allocation, or non-affine index strings---such as symbolic register multiplication, division, or modulo arithmetic---conventional analyses encounter limitations. Because legacy frameworks over-approximate systematically at the first sign of non-linear transformations, they trigger a safe but conservative fallback. This forces them to assume a dependency between all memory statements, resulting in a monolithic heap allocation that destroys potential parallelism. While conservative serialization remains a critical mathematical necessity to guarantee absolute safety when non-linear integer arithmetic cannot be resolved, executing this fallback creates an artificial bottleneck. This lack of precise spatial alias analysis ultimately forces HLS compilers to serialize operations or inject costly multiplexer-driven arbitration and stall logic that degrades target hardware efficiency.

\subsection{A Spatial Resolution: Separation Logic}
To overcome the non-affine bottleneck, we introduce a spatial resolution framework built upon Separation Logic \cite{reynolds2002separation}, an extension of Hoare logic \cite{hoare1969axiomatic} designed for local reasoning about shared mutable data structures \cite{o2004resources}. By utilizing the core separating conjunction operator ($*$), separation logic inherently embeds pointer non-aliasing properties directly into its algebraic structure, asserting that distinct spatial descriptions govern completely disjoint regions of the memory heap. 

While prior hardware synthesis and dependency analysis methods have applied separation logic to source-level C/C++ abstract syntax trees (ASTs) \cite{winterstein2015separation}, our framework operates directly on the LLVM Intermediate Representation (IR), utilizing symbolic execution paradigms established in software verification frameworks like Heapster \cite{he2021type} and llStar \cite{villard2013here}. 
This low-level approach enables compositionality  and allows us to intercept flat address arithmetic computed by the compiler's \texttt{getelementptr} instructions \cite{villard2013here}. 
By maintaining physical memory allocations and pure arithmetic states as distinct subsets, structural bank collisions evaluate to a spatial contradiction ($P \ast P \Rightarrow \perp$),  reducing the memory conflict problem to an algebraic evaluation handled by downstream Satisfiability Modulo Theories (SMT) engines such as Z3 \cite{de2008z3}.

\subsection{Contributions}
In this paper, we formalize the detection of structural memory hazards in HLS scheduling loops by transforming low-level data dependency tracking into spatial logic validation properties. Our core theoretical and architectural contributions include:
\begin{itemize}
    \item \textbf{Formal Hardware Memory Model:} We establish a static memory framework over the LLVM IR, mechanically abstracting multi-dimensional array index expressions via \texttt{getelementptr} instructions and binding them to polymorphic spatial memory bank predicates governing cyclic, block, and complete partitioning transformations.
    \item \textbf{Conflict-Free Unrolling (CFU) Condition:} We formulate a definitive spatial execution requirement, utilizing an iterated separating conjunction over target address offsets that detects physical bank hazards through automatic spatial logic collapse ($\cfu(E,A,N) \Rightarrow \perp$).
    \item \textbf{Automated SMT Translation and Fallback:} We implement a mathematically sound translation function ($\mathcal{T}$) reducing spatial propositions into a pairwise inequality matrix across local Static Single Assignment (SSA) registers, avoiding rigid polyhedral constraints while introducing a conservative sequential fallback model to ensure absolute safety in the presence of undecidable non-linear integer arithmetic expressions.
\end{itemize}

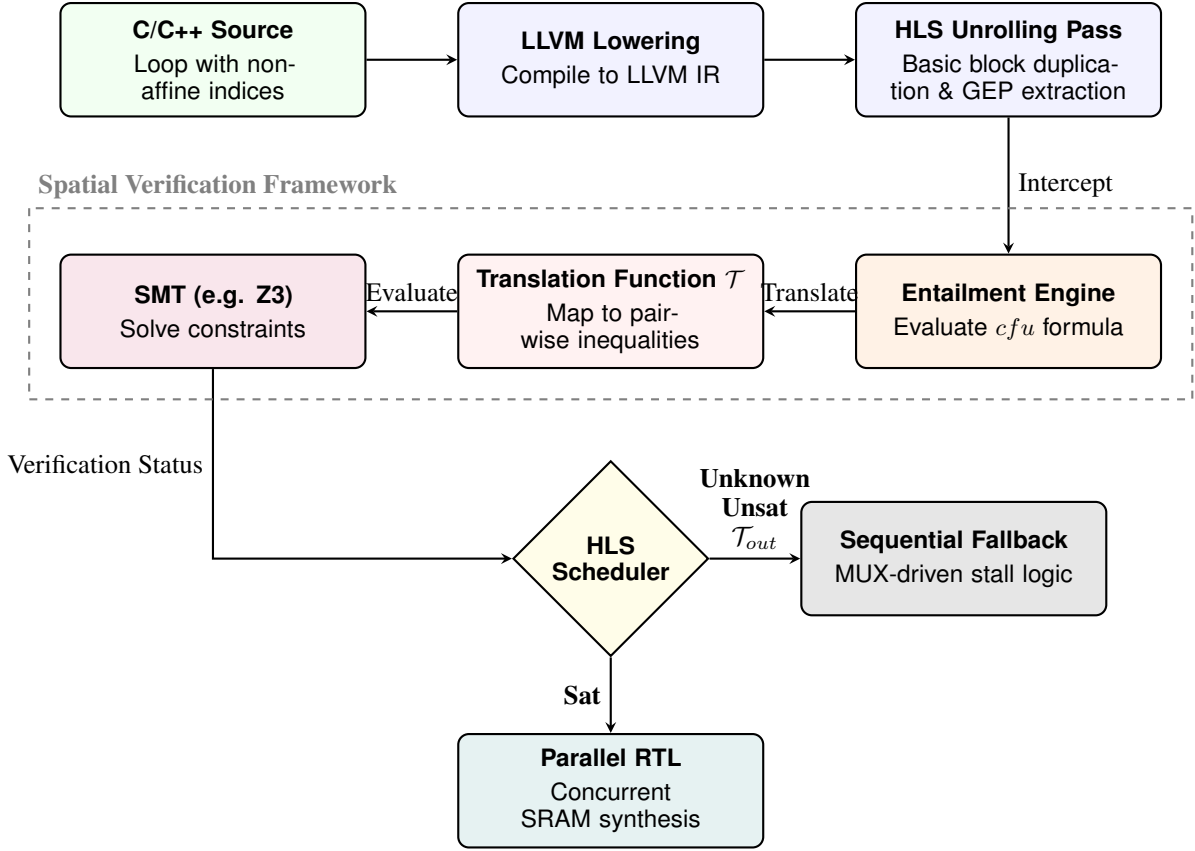
\begin{figure}[htbp]
\centering
\begin{tikzpicture}[
    auto,
    block/.style = {
        rectangle, 
        draw=black, 
        thick,
        fill=blue!5, 
        text width=3.8cm, 
        align=center, 
        rounded corners=4pt, 
        minimum height=1.5cm,
        font=\sffamily\small
    },
    decision/.style = {
        diamond, 
        draw=black, 
        thick,
        fill=yellow!10, 
        text width=2.0cm, 
        align=center, 
        font=\sffamily\small,
        inner sep=0pt
    },
    arrow/.style = {
        thick,
        ->,
        >=stealth
    }
]

    % --- Row 1: HLS Frontend ---
    \node [block, fill=green!5] (source) {
        \textbf{C/C++ Source}\\
        \vspace{0.1cm}
        Loop with non-affine indices
    };

    \node [block, right=1.2cm of source] (lower) {
        \textbf{LLVM Lowering}\\
        \vspace{0.1cm}
        Compile to LLVM IR
    };

    \node [block, right=1.2cm of lower] (unroll) {
        \textbf{HLS Unrolling Pass}\\
        \vspace{0.1cm}
        Basic block duplication \& GEP extraction
    };

    % --- Row 2: Spatial Verification Framework ---
    \node [block, fill=orange!10, below=1.8cm of unroll] (entailment) {
        \textbf{Entailment Engine}\\
        \vspace{0.1cm}
        Evaluate $cfu$ formula
    };

    \node [block, fill=red!5, left=1.2cm of entailment] (translation) {
        \textbf{Translation Function $\mathcal{T}$}\\
        \vspace{0.1cm}
        Map to pairwise inequalities
    };

    \node [block, fill=purple!10, left=1.2cm of translation] (smt) {
        \textbf{SMT (e.g. Z3)}\\
        \vspace{0.1cm}
        Solve constraints
    };

    % --- Row 3: Scheduler & Fallback ---
    \node [decision, below=1.2cm of translation] (scheduler) {
        \textbf{HLS}\\
        \textbf{Scheduler}
    };

    \node [block, fill=gray!20, right=1.2cm of scheduler] (fallback) {
        \textbf{Sequential Fallback}\\
        \vspace{0.1cm}
        MUX-driven stall logic
    };

    % --- Row 4: Parallel RTL ---
    \node [block, fill=teal!10, below=1.0cm of scheduler] (parallel) {
        \textbf{Parallel RTL}\\
        \vspace{0.1cm}
        Concurrent SRAM synthesis
    };

    % --- Arrows ---
    \draw [arrow] (source) -- (lower);
    \draw [arrow] (lower) -- (unroll);
    
    \draw [arrow] (unroll) -- node[right] {Intercept} (entailment);
    \draw [arrow] (entailment) -- node[above] {Translate} (translation);
    \draw [arrow] (translation) -- node[above] {Evaluate} (smt);
    
    % Route SMT status down and across to the scheduler
    \draw [arrow] (smt.south) |- node[near start, left] {Verification Status} (scheduler.west);
    
    % Scheduler branching
    \draw [arrow] (scheduler.south) -- node[left] {\textbf{Sat}} (parallel.north);
    \draw [arrow] (scheduler.east) -- node[above, align=center] {\textbf{Unknown} 
    \\ \textbf{Unsat}
    \\ \textbf{$\mathcal{T}_{out}$}} (fallback.west);

    % --- Bounding Box / Grouping ---
    \draw[dashed, gray, thick] 
        ($(smt.north west) + (-0.4, 0.6)$) rectangle ($(entailment.south east) + (0.4, -0.4)$);
    \node[gray, anchor=south west] at ($(smt.north west) + (-0.4, 0.6)$) {\textbf{Spatial Verification Framework}};

\end{tikzpicture}
\caption{Operational Pipeline: The Spatial Verification Framework informs the HLS Scheduler.}
\label{fig:active_scheduling_pipeline}
\end{figure}

\section{Background: Arrays, Loops, and Hyperplanes}

\subsection{Array Partitioning and Loop Unrolling in HLS}
When a High-Level Synthesis (HLS) compiler applies loop unrolling by a factor of $U$, it performs standard basic block duplication to expose instruction-level parallelism. This optimization dictates that $U$ concurrent memory accesses must be safely scheduled within a single hardware clock cycle to maximize throughput. However, synthesized on-chip SRAMs typically feature a limited number of access ports. To sustain the required parallel bandwidth, HLS tools employ array partitioning directives that mechanically divide a monolithic array footprint into $N$ distinct physical memory banks. Common strategies include cyclic partitioning, where consecutive elements are interleaved round-robin across banks at offsets $j \cdot N + k$, and block partitioning, which divides the array into contiguous chunks starting at $\alpha + k \cdot B$. To avoid structural hazards—which occur when concurrent operations attempt to access the same single-ported memory bank simultaneously—the scheduler must mathematically guarantee that the target addresses map to strictly independent physical banks.

\subsection{The Non-Affine Bottleneck in Polyhedral Analysis}
To prove that duplicated memory accesses target independent banks, standard HLS compilers historically rely on the polyhedral model. This framework maps loop iteration spaces into multi-dimensional geometric hyperplanes to perform exact dependency analysis. However, polyhedral analysis strictly requires loop bounds and memory access functions to be affine—that is, linear combinations of enclosing loop variables and constants.

When array indices involve non-affine arithmetic—such as arbitrary symbolic variable multiplication, data-dependent indexing, or the modulo and division operations fundamentally required to evaluate cyclic and block bank mapping functions—the geometric model breaks down. Pioneering extensions of the polyhedral framework have successfully mitigated some of these limitations by elegantly employing parametric analysis to synthesize lightweight runtime checks and dynamic pipeline breaks, successfully parallelizing loops with uncertain or nonuniform dependencies \cite{liu2017polyhedral}. 
These advanced dynamic techniques provide a robust foundation for runtime hazard resolution; however, to remain tractable within standard integer set libraries, they typically require the underlying dependence distances to retain an affine structure, and they can introduce specialized runtime controller logic to manage pipeline execution speeds dynamically \cite{liu2017polyhedral}. 
To complement these runtime approaches, our framework aims to address non-linear address arithmetic statically at compile-time. 
By utilizing spatial logic directly over flattened intermediate expressions, we look to resolve complex bank mapping dependencies without relying on linear geometric bounds or generating additional runtime state-machine overhead.

\subsection{Separation Logic in Hardware Verification}
Separation Logic (SL) offers a rigorous mathematical foundation to reason about spatially distributed memory states, originally designed for shared mutable data structures \cite{reynolds2002separation, o2004resources}. The fundamental unit of spatial resource isolation in SL is the separating conjunction operator ($\ast$). The spatial formula $\Sigma_1 \ast \Sigma_2$ asserts that the memory heap can be partitioned into strictly disjoint hardware regions, embedding the non-aliasing property of pointers directly into the algebraic framework. 
A powerful consequence of this strict disjointness is that a single spatial resource cannot be exclusively owned by multiple concurrent execution entities; evaluating $P \ast P$ instantly yields a logical contradiction ($P \ast P \Rightarrow \perp$) \cite{berdine2005symbolic,brookes2016concurrent}. 

Recent advancements have successfully applied SL to hardware verification and HLS parallelization \cite{winterstein2015separation}. 
Existing SL-based HLS tools focus on dynamic memory allocation, pointer-chasing, and shape analysis of recursive dynamic data structures (e.g., trees and linked lists) by inferring complex fix-point loop invariants. 
However, static array partitioning and flat loop unrolling still need to be studied. 
Our framework explicitly bridges this gap, adapting the disjointness axioms of SL down to the LLVM Intermediate Representation (IR) level \cite{he2021type,villard2013here} to natively evaluate partitioned array bank mapping expressions, bypassing the polyhedral affine bottleneck.

\section{HLS Memory Model in Separation Logic}

\paragraph{Formal Syntax of the Spatial Framework}
To ground the interactions between programmatic variables, symbolic non-affine expressions, and physical memory bank assets, we formally define the core language syntax via the following Backus-Naur Form (BNF) grammar. Unlike traditional software-oriented Separation Logic \cite{reynolds2002separation,o2004resources}, which operates over abstract syntax trees (ASTs), our framework operates directly on the LLVM Intermediate Representation (IR), following the approach of \cite{he2021type,villard2013here}.

Let $\text{Reg}$ be the set of immutable LLVM virtual registers and global identifiers, $\text{Reg}'$ be the set of auxiliary primed variables, and $\text{Const}$ be the set of constant natural numbers. We distinguish a specific subset $\text{ArrID} \subset \text{Reg}$ to represent the universe of synthesized array identifiers (e.g., $@A, $@B).

\begin{align*}
e \in \text{Expr} \quad &::= \quad c \in \text{Const} \mid x \in \text{Reg} \mid x' \in \text{Reg}' \mid e_1 + e_2 \mid e_1 - e_2 \mid e_1 \cdot e_2 \mid e_1 \pmod{e_2} \mid \lfloor e_1 / e_2 \rfloor \\
\Pi \in \text{Pure} \quad &::= \quad \text{true} \mid \text{false} \mid e_1 = e_2 \mid e_1 \neq e_2 \mid \Pi_1 \land \Pi_2 \\
\Sigma \in \text{Spatial} \quad &::= \quad \text{emp} \mid e_1 \mapsto e_2 \mid \bank(e_1, A, e_2) \mid \arr(A) \mid \Sigma_1 * \Sigma_2 \quad (\text{where } A \in \text{ArrID})
\end{align*}

Where $e_1 \cdot e_2$ natively models arbitrary symbolic variable multiplication, allowing non-affine index strings to step into the spatial heap definition before being resolved by a downstream SMT engine. The parameters within $\bank(e_k, A, e_N)$ follow the expression grammar rule ($e$), proving that bank locations are fundamentally treated as dynamic algebraic variants rather than rigid static integers.

\paragraph{Mechanical 1D Flattening via \texttt{getelementptr}}
By shifting the semantic model to the LLVM IR level, our framework fundamentally bypasses the need to mathematically simulate the row-major flattening of multi-dimensional C/C++ arrays. The LLVM compiler mechanically lowers all complex array indexing into flat 1D arithmetic via the \texttt{getelementptr} (GEP) instruction. Consequently, the spatial framework simply extracts the resulting flat arithmetic expression $e$ from the GEP instruction and feeds it directly into the bank mapping function, drastically simplifying the required operational semantics.

\subsection{The Spatial Memory Heap and Store}

To mathematically reason about High-Level Synthesis (HLS) architectures, we adapt the foundational framework of Separation Logic to model program execution and physical SRAM instances as distinct state components. Separation logic formally describes the program state using two distinct elements: the store and the heap.

\paragraph{The Monotonic Store ($s$)}
Because LLVM IR strictly enforces Static Single Assignment (SSA), our framework  avoids the state-space explosion associated with modeling mutable C/C++ variables. 
Let $\text{Reg}$ denote the set of LLVM virtual registers and $\text{Val}$ represent the set of storable hardware values (which includes the set of physical memory addresses, $\mathbb{L}$). The store mathematically captures the immediate, immutable state of local registers in the synthesized hardware. It is defined as a partial function mapping registers to their currently assigned values:
$$s : \text{Reg} \rightharpoonup \text{Val}$$

Due to SSA, this mapping is strictly monotonic during the symbolic execution of a basic block; virtual registers are never overwritten. For example, an evaluation of the store where $s(\%x) = 3$ indicates that the virtual register $\%x$ is permanently bound to the scalar value $3$ for that execution path, eliminating the need for state-update tracking ($s \to s'$).

\paragraph{The Spatial Heap ($h$)}
While the store resolves register values, the heap models the spatially distributed on-chip memory (e.g., block RAMs). The heap, $h$, is a finite partial function mapping physical, addressable memory locations to hardware values:
$$h : \mathbb{L} \rightharpoonup \text{Val}$$

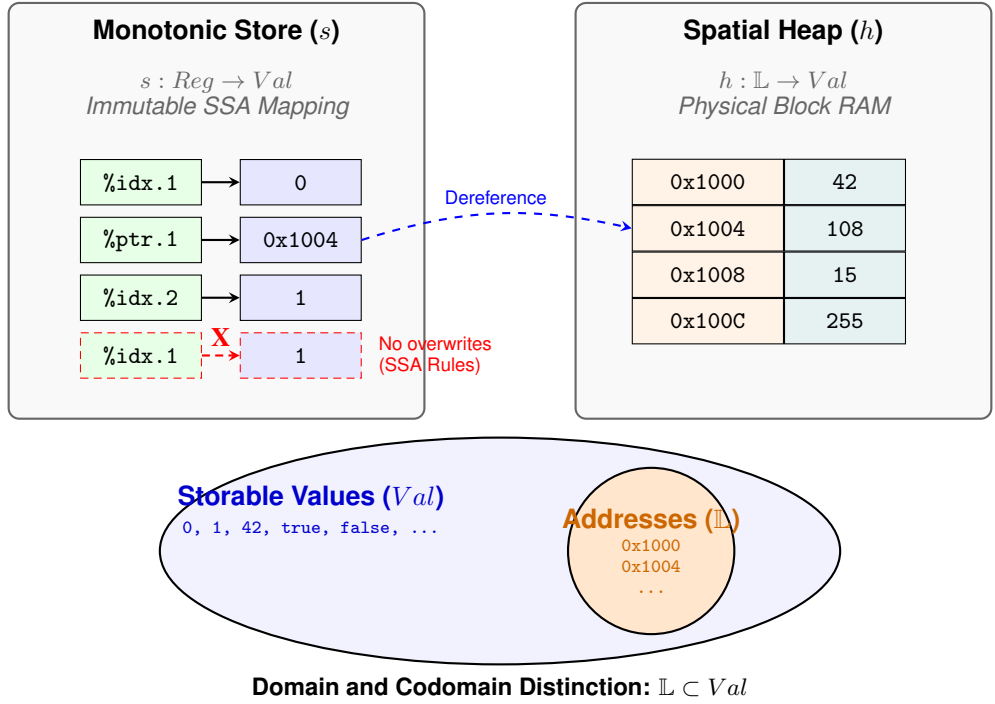
\begin{figure}[htbp]
\centering
\begin{tikzpicture}[
    auto,
    >=stealth,
    panel/.style = {
        rectangle,
        draw=black!60,
        thick,
        rounded corners=4pt,
        fill=gray!5,
        minimum width=5.5cm,
        minimum height=5.5cm
    },
    title/.style = {
        font=\sffamily\bfseries,
        align=center
    },
    subtitle/.style = {
        font=\sffamily\small\itshape,
        align=center,
        text=gray!80!black
    },
    reg/.style = {
        rectangle,
        draw=black,
        fill=green!10,
        minimum width=1.6cm,
        minimum height=0.6cm,
        font=\ttfamily\small
    },
    val/.style = {
        rectangle,
        draw=black,
        fill=blue!10,
        minimum width=1.6cm,
        minimum height=0.6cm,
        font=\ttfamily\small
    },
    memaddr/.style = {
        rectangle,
        draw=black,
        fill=orange!10,
        minimum width=2cm,
        minimum height=0.6cm,
        font=\ttfamily\small
    }
]

    % ---------------------------------------------------
    % PANELS (Drawn first to act as the background layer)
    % ---------------------------------------------------
    \node[panel] (store_panel) at (0,0) {};
    \node[panel] (heap_panel) at (7.5,0) {};

    % ---------------------------------------------------
    % LEFT PANEL: Monotonic Store Nodes
    % ---------------------------------------------------
    \node[title] (store_title) at ([yshift=-0.4cm]store_panel.north) {Monotonic Store ($s$)};
    \node[subtitle, below=0.1cm of store_title] (store_sub) {$s: Reg \to Val$\\Immutable SSA Mapping};
    
    \node[reg, below=0.4cm of store_sub, xshift=-1cm] (reg1) {\%idx.1};
    \node[val, right=0.5cm of reg1] (val1) {0};
    \draw[->, thick] (reg1) -- (val1);
    
    \node[reg, below=0.15cm of reg1] (reg2) {\%ptr.1};
    \node[val, right=0.5cm of reg2] (val2) {0x1004};
    \draw[->, thick] (reg2) -- (val2);

    \node[reg, below=0.15cm of reg2] (reg3) {\%idx.2};
    \node[val, right=0.5cm of reg3] (val3) {1};
    \draw[->, thick] (reg3) -- (val3);

    % SSA overwrite crossed out
    \node[reg, draw=red, densely dashed, below=0.15cm of reg3] (reg4) {\%idx.1};
    \node[val, draw=red, densely dashed, right=0.5cm of reg4] (val4) {1};
    \draw[->, thick, red, densely dashed] (reg4) -- (val4) node[midway, sloped, red] {\textbf{X}};
    \node[right=0.1cm of val4, text=red, font=\sffamily\scriptsize, align=left] {No overwrites\\(SSA Rules)};

    % ---------------------------------------------------
    % RIGHT PANEL: Spatial Heap Nodes
    % ---------------------------------------------------
    \node[title] (heap_title) at ([yshift=-0.4cm]heap_panel.north) {Spatial Heap ($h$)};
    \node[subtitle, below=0.1cm of heap_title] (heap_sub) {$h: \mathbb{L} \to Val$\\Physical Block RAM};

    % Memory Array (BRAM)
    \node[memaddr, below=0.4cm of heap_sub, xshift=-1cm] (addr1) {0x1000};
    \node[val, fill=teal!10, right=0cm of addr1] (memval1) {42};
    
    \node[memaddr, below=0cm of addr1] (addr2) {0x1004};
    \node[val, fill=teal!10, right=0cm of addr2] (memval2) {108};
    
    \node[memaddr, below=0cm of addr2] (addr3) {0x1008};
    \node[val, fill=teal!10, right=0cm of addr3] (memval3) {15};
    
    \node[memaddr, below=0cm of addr3] (addr4) {0x100C};
    \node[val, fill=teal!10, right=0cm of addr4] (memval4) {255};

    % Arrow showing ptr mapping to physical memory
    \draw[->, thick, blue, dashed] (val2.east) to[bend left=15] node[above, font=\sffamily\scriptsize] {Dereference} (addr2.west);

    % ---------------------------------------------------
    % BOTTOM PANEL: Venn Diagram
    % ---------------------------------------------------
    % Placed explicitly below the panels
    \coordinate (center_bottom) at (3.75, -4.5);
    
    % Draw Val set (larger ellipse)
    \draw[thick, fill=blue!5] (center_bottom) ellipse (4.5cm and 1.5cm);
    \node[font=\sffamily\bfseries, text=blue!80!black] at ($(center_bottom) + (-2.5, 0.7)$) {Storable Values ($Val$)};
    \node[font=\ttfamily\scriptsize, text=blue!80!black] at ($(center_bottom) + (-2.5, 0.3)$) {0, 1, 42, true, false, ...};

    % Draw L set (smaller circle inside)
    \draw[thick, fill=orange!20] ($(center_bottom) + (2.0, 0)$) circle (1.1cm);
    \node[font=\sffamily\bfseries, text=orange!80!black] at ($(center_bottom) + (2.0, 0.4)$) {Addresses ($\mathbb{L}$)};
    \node[font=\ttfamily\scriptsize, text=orange!80!black, align=center] at ($(center_bottom) + (2.0, -0.2)$) {0x1000\\0x1004\\...};
    
    % Relation text
    \node[font=\sffamily\small\bfseries, align=center, fill=white, inner sep=2pt] at ($(center_bottom) + (0, -1.8)$) {Domain and Codomain Distinction: $\mathbb{L} \subset Val$};

\end{tikzpicture}
\caption{Dual-State Architecture: Mathematically isolating pure arithmetic states from spatial memory allocations. The Monotonic Store ($s$) binds SSA virtual registers to storable values, while the Spatial Heap ($h$) maps physical address boundaries to memory contents. Crucially, physical addresses ($\mathbb{L}$) form a strict subset of all storable values ($Val$).}
\label{fig:dual_state}
\end{figure}

\paragraph{Domain and Codomain Distinction ($\mathbb{L} \subset \text{Val}$)}
It is a critical mathematical distinction of this framework that the codomain of the store ($\text{Val}$) and the domain of the heap ($\mathbb{L}$) are strictly not equivalent. In synthesized hardware, all memory addresses are values (typically unsigned bitvectors), but not all values (e.g., floating-point results, loop counters, booleans) are valid memory addresses. Therefore, the set of memory locations is a strict subset of all storable values: 
$$\mathbb{L} \subset \text{Val}$$

This intentional mismatch dictates the operational semantics of pointer dereferencing. When evaluating a spatial assertion such as $\%x \mapsto v$, the state routes data through both functions sequentially: first, the store evaluates the register pointer $s(\%x) = l$ (where $l \in \mathbb{L}$); second, the heap retrieves the data $h(l) = v$ (where $v \in \text{Val}$). By maintaining $\mathbb{L}$ and $\text{Val}$ as distinct sets, the theoretical framework mathematically isolates pure arithmetic states from spatial memory allocations.

\paragraph{Spatial Ownership and Disjointness}
The fundamental unit of spatial resource in this framework is the points-to relation. The assertion $E \mapsto F$ strictly dictates that a physical SRAM address evaluated from expression $E$ currently stores the value evaluated from $F$, asserting exclusive ownership over that single physical location. 

In physical hardware, structural hazards occur when concurrent parallel operations attempt to access the same single-ported memory bank simultaneously. We abstract the absence of such hazards through the strict mathematical property of heap disjointness. Two hardware memory heaps, $h_1$ and $h_2$, are strictly disjoint, denoted by $h_1 \perp h_2$, if and only if their respective domains share no common physical addresses:
$$h_1 \perp h_2 \iff \text{dom}(h_1) \cap \text{dom}(h_2) = \emptyset$$

\paragraph{The Separating Conjunction}
This disjointness is elegantly encapsulated by the separating conjunction operator ($*$). The spatial formula $\Sigma_1 * \Sigma_2$ asserts that the current heap can be partitioned into two strictly disjoint hardware regions (heaplets), $h_1$ and $h_2$, such that $\Sigma_1$ holds for $h_1$ and $\Sigma_2$ holds for $h_2$. 

This operator embeds the non-aliasing property of pointers directly into the algebraic framework (e.g., $E \mapsto F_1 * G \mapsto F_2 \implies E \neq G$). By reducing parallel structural safety to spatial disjointness, HLS compilers can synthesize parallel memory accesses without conservative approximations of non-affine dependencies. Finally, the empty heap assertion, $\text{emp}$, denotes a state where the active execution claims ownership of no physical memory resources.

\subsection{Memory Bank Predicates}

In High-Level Synthesis (HLS) flows built upon modern compiler infrastructure, source-level partitioning directives are lowered into LLVM IR metadata. This metadata explicitly instructs the backend to partition monolithic arrays into $N$ distinct physical memory banks to increase memory bandwidth and enable parallel data access. To model this hardware transformation mathematically at the bitcode level, we introduce hardware-aware spatial predicates that encapsulate the exclusive ownership of specific partitioned memory banks.

Let array $A$ be a finite contiguous allocation of $M$ elements starting at physical base address $\alpha \in \mathbb{L}$. We define the predicate $\bank(k, A, N)$ to assert exclusive spatial ownership over the $k$-th physical memory bank, where $0 \le k < N$. The precise physical addresses governed by this predicate depend strictly on the HLS partitioning strategy employed.

\paragraph{The Static Hardware Environment ($\Gamma$)}
In our syntax, $A \in \text{ArrID}$ acts purely as a symbolic identifier tied to an LLVM virtual register. To mathematically derive the physical constraints of the array---such as its base address and bounds---we introduce a fixed static hardware environment, $\Gamma$. This environment acts as the formal equivalent of the LLVM module's global symbol table and bitcode metadata, mapping an array identifier to its fixed hardware metadata tuple: its physical base address ($\alpha \in \mathbb{L}$), its total element capacity ($M \in \mathbb{N}$), and its partitioning scheme ($\mathcal{P}$).

$$ \Gamma : \text{ArrID} \to \mathbb{L} \times \mathbb{N} \times \{ \text{Cyclic}, \text{Block}, \text{Complete} \} $$

For any array $A$, its metadata is formally extracted via:
$$ \Gamma(A) = (\alpha, M, \mathcal{P}) $$

Consequently, when the execution state evaluates a spatial predicate such as $\bank(k, A, N)$, the logical framework queries $\Gamma(A)$ to anchor the abstract identifier $A$ to the concrete physical SRAM address space defined by $\alpha$ and bounded by $M$.

\paragraph{Cyclic Partitioning}
Under cyclic partitioning, consecutive array elements are interleaved across the $N$ memory banks in a round-robin fashion. The $k$-th bank contains elements at offsets $j \cdot N + k$. We formalize the spatial ownership of a cyclic bank using the iterated separating conjunction ($\bigast$):
$$\bank_{cyc}(k, A, N) \triangleq \bigast_{j=0}^{\lfloor (M-1-k)/N \rfloor} (\alpha + j \cdot N + k \mapsto v_{j,k})$$
where $v_{j,k} \in \text{Val}$ represents the value stored at that specific 1D SRAM address offset.

\paragraph{Block Partitioning}
Under block partitioning, the array is divided into $N$ contiguous chunks. Let $B = \lceil M / N \rceil$ denote the block size. The $k$-th bank owns a contiguous sequence of addresses starting at $\alpha + k \cdot B$. The spatial ownership for a block bank is formalized as:
$$\bank_{blk}(k, A, N) \triangleq \bigast_{j=0}^{\min(B-1, M - 1 - k \cdot B)} (\alpha + k \cdot B + j \mapsto v_{j,k})$$

\paragraph{Complete Partitioning (The Boundary Case)}
The third standard HLS directive is complete partitioning, which dissolves the SRAM entirely into discrete, independent hardware registers. Mathematically, this is modeled as a strict boundary case of either block or cyclic partitioning where the number of partitioned banks equals the total number of array elements ($N = M$). Consequently, the bank size is exactly 1, and the iterated conjunction collapses into a singular spatial points-to assertion for each physical index:
$$\bank_{comp}(k, A, M) \triangleq \alpha + k \mapsto v_k$$
This demonstrates that our spatial framework universally captures all HLS memory architectures, scaling seamlessly from monolithic SRAMs ($N=1$) down to fully unrolled register files ($N=M$).

\paragraph{The Polymorphic Bank Predicate}
To maintain the universality of our theoretical framework across diverse hardware generation strategies, we define the generic spatial predicate $\bank(k, A, N)$ as a polymorphic interface. During the static symbolic execution of an LLVM basic block, this abstract predicate is dynamically instantiated into a concrete spatial representation based on the specific HLS partitioning metadata attached to the array's allocation instruction in the IR.

Let $\mathcal{P} \in \{\text{Cyclic}, \text{Block}, \text{Complete}\}$ denote the specific partitioning scheme applied to array $A$. The abstract bank predicate is formally defined as a piecewise mapping:
$$
\bank(k, A, N) \triangleq 
\begin{cases} 
\bank_{cyc}(k, A, N) & \text{if } \mathcal{P} = \text{Cyclic} \\
\bank_{blk}(k, A, N) & \text{if } \mathcal{P} = \text{Block} \\
\bank_{comp}(k, A, N) & \text{if } \mathcal{P} = \text{Complete}
\end{cases}
$$

Crucially, the parameters $k$ and $N$ are not restricted to compile-time natural constants; they are treated as symbolic expressions evaluated under the SSA store $s$ and path condition $\pi$, thereby natively accommodating non-affine variable terms (e.g., $k = (x \cdot Y + y) \pmod N$) whose functional disjointness is deferred to algebraic evaluation.

This mathematical abstraction is critical for the robustness of the framework. It guarantees that our core theorems remain universally valid for any partitioned array. Furthermore, it explicitly links the high-level Separation Logic axioms directly to the LLVM IR, allowing the Entailment Engine to apply the exact algebraic mapping function (e.g., modulo for cyclic, division for block) required by the SMT solver to verify physical disjointness.

\paragraph{The Array Composition Rule}
Regardless of the partitioning strategy, the fundamental physical reality of the synthesized hardware is that the complete array is merely the disjoint spatial union of its constituent banks. We prove this mathematically via the array composition rule. 

A complete array predicate, $\arr(A)$, is strictly equivalent to the iterated separating conjunction of all $N$ disjoint bank predicates:
$$\arr(A) \triangleq \bigast_{k=0}^{N-1} \bank(k, A, N)$$

This formulation is mathematically powerful because it inherently guarantees structural non-interference. By the fundamental axioms of Separation Logic, the iterated separating conjunction strictly enforces that for any two distinct banks $k_1 \neq k_2$, their physical address domains are completely disjoint ($\text{dom}(\bank_{k_1}) \cap \text{dom}(\bank_{k_2}) = \emptyset$). 

During the symbolic execution of a duplicated LLVM basic block (resulting from a loop unroll pass), \texttt{getelementptr} instructions compute flat 1D address offsets. If two concurrent \texttt{load} or \texttt{store} instructions subsequently attempt to claim ownership of the same physical bank $k$ using these offsets, the execution state will attempt to evaluate $\bank(k) * \bank(k)$. According to the core properties of the separating conjunction, $P * P \implies \bot$ (a logical contradiction). This logical contradiction mathematically flags the exact presence of a structural memory conflict, entirely bypassing the need for affine constraint resolution.

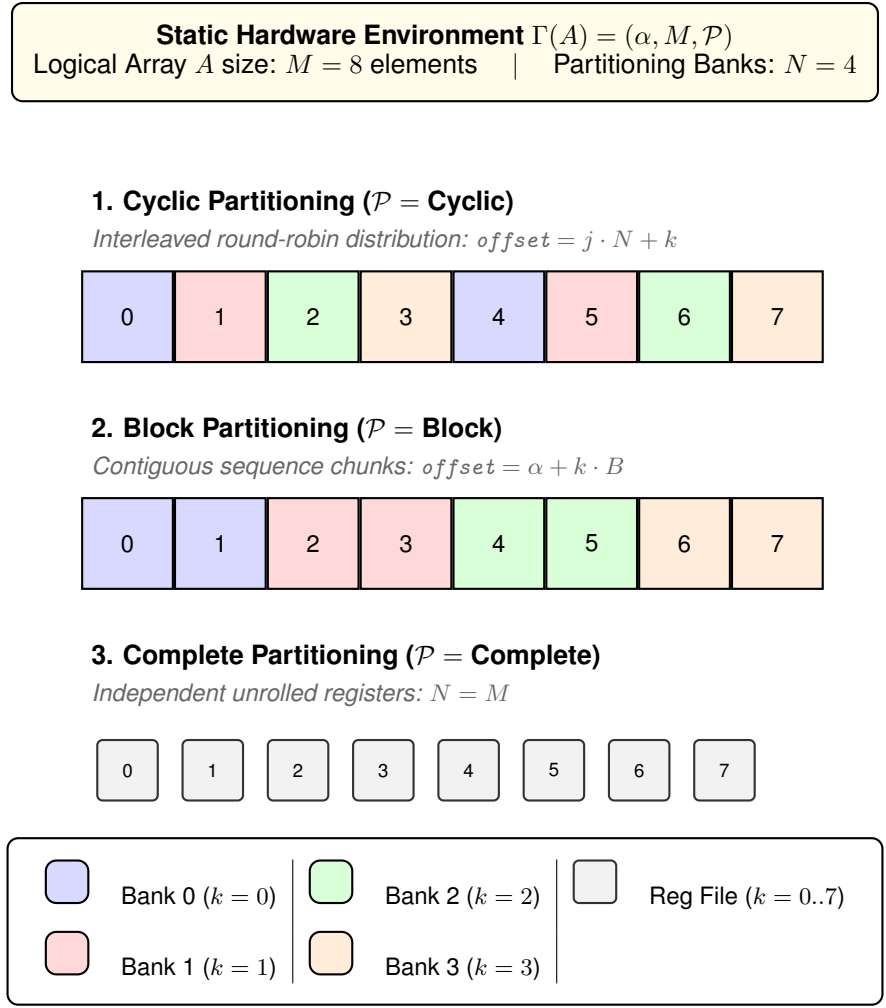
\begin{figure}[htbp]
\centering
\begin{tikzpicture}[
    auto,
    >=stealth,
    cell/.style={
        rectangle, 
        draw=black, 
        thick, 
        minimum width=1.2cm, 
        minimum height=1.2cm, 
        align=center, 
        font=\sffamily\small
    },
    reg/.style={
        rectangle, 
        draw=black!80, 
        thick, 
        minimum width=0.8cm, 
        minimum height=0.8cm, 
        align=center, 
        font=\sffamily\scriptsize,
        rounded corners=2pt,
        fill=gray!10
    },
    b0/.style={fill=blue!15},
    b1/.style={fill=red!15},
    b2/.style={fill=green!15},
    b3/.style={fill=orange!15},
    title/.style={
        font=\sffamily\bfseries,
        anchor=west
    },
    desc/.style={
        font=\sffamily\footnotesize\itshape,
        anchor=west,
        text=gray!80!black
    }
]

    % --- Global Metadata (Centered at x=4.8 to align with arrays) ---
    \node[draw=black, thick, fill=yellow!10, rounded corners=4pt, inner sep=8pt, align=center, font=\sffamily] (metadata) at (4.8, 1.5) {
        \textbf{Static Hardware Environment $\Gamma(A) = (\alpha, M, \mathcal{P})$}\\
        Logical Array $A$ size: $M=8$ elements \quad $|\quad$ Partitioning Banks: $N=4$
    };

    % --- 1. Cyclic Partitioning ---
    \node[title] (cyc_title) at (0, -0.5) {1. Cyclic Partitioning ($\mathcal{P} = \text{Cyclic}$)};
    \node[desc] (cyc_desc) at (0, -1) {Interleaved round-robin distribution: \texttt{offset} $= j \cdot N + k$};
    
    \node[cell, b0] (c0) at (0.6, -2) {0};
    \node[cell, b1] (c1) [right=0cm of c0] {1};
    \node[cell, b2] (c2) [right=0cm of c1] {2};
    \node[cell, b3] (c3) [right=0cm of c2] {3};
    \node[cell, b0] (c4) [right=0cm of c3] {4};
    \node[cell, b1] (c5) [right=0cm of c4] {5};
    \node[cell, b2] (c6) [right=0cm of c5] {6};
    \node[cell, b3] (c7) [right=0cm of c6] {7};

    % --- 2. Block Partitioning ---
    \node[title] (blk_title) at (0, -3.5) {2. Block Partitioning ($\mathcal{P} = \text{Block}$)};
    \node[desc] (blk_desc) at (0, -4) {Contiguous sequence chunks: \texttt{offset} $= \alpha + k \cdot B$};
    
    \node[cell, b0] (b0) at (0.6, -5) {0};
    \node[cell, b0] (b1) [right=0cm of b0] {1};
    \node[cell, b1] (b2) [right=0cm of b1] {2};
    \node[cell, b1] (b3) [right=0cm of b2] {3};
    \node[cell, b2] (b4) [right=0cm of b3] {4};
    \node[cell, b2] (b5) [right=0cm of b4] {5};
    \node[cell, b3] (b6) [right=0cm of b5] {6};
    \node[cell, b3] (b7) [right=0cm of b6] {7};

    % --- 3. Complete Partitioning ---
    \node[title] (comp_title) at (0, -6.5) {3. Complete Partitioning ($\mathcal{P} = \text{Complete}$)};
    \node[desc] (comp_desc) at (0, -7) {Independent unrolled registers: $N = M$};
    
    \node[reg] (r0) at (0.6, -8) {0};
    \node[reg] (r1) [right=0.3cm of r0] {1};
    \node[reg] (r2) [right=0.3cm of r1] {2};
    \node[reg] (r3) [right=0.3cm of r2] {3};
    \node[reg] (r4) [right=0.3cm of r3] {4};
    \node[reg] (r5) [right=0.3cm of r4] {5};
    \node[reg] (r6) [right=0.3cm of r5] {6};
    \node[reg] (r7) [right=0.3cm of r6] {7};

    % --- Legend (Moved to bottom center for clean formatting) ---
    \node[draw=black, thick, fill=white, inner sep=8pt, rounded corners=4pt, font=\sffamily\small] (legend) at (4.8, -10) {
        \begin{tabular}{cl | cl | cl}
            \tikz{\node[cell, b0, minimum width=0.4cm, minimum height=0.4cm] {};} & Bank 0 ($k=0$) &
            \tikz{\node[cell, b2, minimum width=0.4cm, minimum height=0.4cm] {};} & Bank 2 ($k=2$) &
            \tikz{\node[reg, minimum width=0.4cm, minimum height=0.4cm] {};} & Reg File ($k=0..7$) \\[1.5ex]
            \tikz{\node[cell, b1, minimum width=0.4cm, minimum height=0.4cm] {};} & Bank 1 ($k=1$) &
            \tikz{\node[cell, b3, minimum width=0.4cm, minimum height=0.4cm] {};} & Bank 3 ($k=3$) & & \\
        \end{tabular}
    };

\end{tikzpicture}
\caption{Polymorphic Bank Predicate Layouts: The abstract memory bank predicate dynamically instantiates into three distinct hardware representations dictated by the static IR metadata $\Gamma$. Cyclic partitioning interleaves elements, Block partitioning maintains contiguous chunks, and Complete partitioning completely dissolves the address space into discrete independent registers.}
\label{fig:poly_bank_predicates}
\end{figure}

\section{Catching Memory Conflicts in Loop Unrolling}

\subsection{The Conflict-Free Unrolling (CFU) Condition}

When a High-Level Synthesis (HLS) compiler applies loop unrolling by a factor of $U$, it does not fundamentally execute a source-level C/C++ AST transformation. Instead, it performs standard LLVM basic block duplication. Each unrolled iteration generates a fresh, immutable set of Static Single Assignment (SSA) virtual registers (e.g., \texttt{\%idx.1}, \texttt{\%idx.2}). This makes the symbolic execution completely linear across the unroll factor $U$. To parallelize these duplicated basic blocks safely within a single hardware clock cycle, the compiler must guarantee that the concurrent execution does not induce structural memory hazards. We formalize the absence of these hazards through the Conflict-Free Unrolling (CFU) condition.

\paragraph{Extraction of Concurrent Memory Accesses}
During the symbolic execution of these unrolled basic blocks, spatial ownership is not triggered by C language level array indexing. Instead, it is natively asserted when the framework intercepts concurrent LLVM \texttt{load} or \texttt{store} instructions targeting a partitioned array $A$. These memory instructions consume flat 1D address offsets computed natively by preceding \texttt{getelementptr} (GEP) instructions. The scheduler extracts this target set of $U$ algebraic GEP expressions. We denote this set of concurrent memory accesses as $E$:
$$E = \{expr_1, expr_2, \dots, expr_U\}$$
where each $expr_i$ represents the pure algebraic evaluation of the flat array offset for the $i$-th parallel iteration. 

To determine the physical hardware routing of these expressions, we define a bank mapping function, $\mathcal{B}(expr_i, N)$, which maps an evaluated GEP arithmetic offset ($expr_i$) to its corresponding physical bank given a partition factor $N$. This mapping is mathematically bound to the array's partitioning scheme $\mathcal{P}$, extracted from the static hardware environment $\Gamma(A) = (\alpha, M, \mathcal{P})$:
$$
\mathcal{B}(expr_i, N) \triangleq 
\begin{cases} 
expr_i \pmod N & \text{if } \mathcal{P} = \text{Cyclic} \\
\lfloor expr_i / B \rfloor & \text{if } \mathcal{P} = \text{Block} \quad (\text{where } B = \lceil M/N \rceil) \\
expr_i & \text{if } \mathcal{P} = \text{Complete}
\end{cases}
$$

\paragraph{Formulation of the CFU Condition}
For the HLS scheduler to safely synthesize these $U$ accesses in parallel, the compile-time symbolic state must demonstrate concurrent and exclusive spatial ownership over the target memory banks in the spatial heap $H$. We model this simultaneous scheduling requirement by demanding that $H$ can satisfy the iterated separating conjunction ($\bigast$) of the required bank predicates.

We formally define the Conflict-Free Unrolling condition for the target set $E$ as follows.
$$\cfu(E, A, N) \triangleq \bigast_{i=1}^{U} \bank(\mathcal{B}(expr_i, N), A, N)$$

The iterated separating conjunction requires the existence of $U$ strictly disjoint heaplets, each satisfying the exclusive spatial ownership of a required physical bank. Because the path condition $\pi$ handles pure LLVM IR algebra—natively accommodating non-affine operations like modulo or division evaluated by the bank mapping function—the framework completely bypasses strict polyhedral geometric models. 

The detection of a memory conflict is thus reduced to an SMT algebraic equivalence query over the SSA variables. If iteration $i$ claims ownership of $\bank(k_1)$ and iteration $j$ claims $\bank(k_2)$, the spatial logic strictly requires $\bank(k_1) * \bank(k_2)$. To enforce spatial disjointness, the framework can  mechanically generate the Verification Condition (VC):
$$\pi \implies k_1 \neq k_2$$

If the SMT oracle evaluates this VC to false or unknown—meaning it cannot mathematically prove the physical indices are distinct—the system explicitly flags a structural memory hazard. 
Consequently, if $\cfu(E, A, N)$ evaluates to a satisfiable spatial state rather than $\bot$, it serves as proof that the synthesized parallel Verilog execution trace is free of memory bank collisions.

\subsection{Spatial Collapse as Conflict Detection}

At the core of Separation Logic lies the principle of exclusive spatial ownership. The separating conjunction, $P * Q$, strictly dictates that the spatial propositions $P$ and $Q$ must hold over disjoint regions of the memory heap. A direct consequence of this foundational axiom is that a single spatial resource cannot be disjointly owned by multiple execution entities simultaneously. For any non-empty spatial predicate $P$, attempting to separate it from itself yields a logical contradiction:
$$P * P \implies \bot$$

We leverage this inherent spatial disjointness as a mathematical mechanism to natively detect structural memory hazards in HLS scheduling. When the Conflict-Free Unrolling (CFU) condition is evaluated, a memory conflict manifests mathematically as a spatial collapse triggered by the Entailment Engine and SMT oracle. 

Consider two concurrent LLVM memory instructions (e.g., \texttt{load} or \texttt{store}) from distinct unrolled basic blocks, $i$ and $j$ (where $1 \le i < j \le U$). Their respective \texttt{getelementptr} instructions compute flat 1D offsets that map to physical memory banks $k_i = \mathcal{B}(expr_i, N)$ and $k_j = \mathcal{B}(expr_j, N)$. To safely schedule these instructions simultaneously, the spatial logic strictly requires $\bank(k_i, A, N) * \bank(k_j, A, N)$.

To mathematically prove this spatial disjointness, the framework generates a Verification Condition (VC) bounded by the pure algebraic path condition $\pi$:
$$\pi \implies k_i \neq k_j$$

If the SMT oracle cannot prove this VC—either evaluating it to ``false'' or returning ``Unknown'' due to the undecidability of non-linear arithmetic embedded in the SSA variables—the framework safely assumes a hazard exists where both accesses target the same physical bank $k$. When constructing the CFU condition for this parallel execution state, the iterated separating conjunction forcefully expands to demand concurrent ownership of bank $k$ twice:
$$\cfu(E, A, N) = \bank(k, A, N) * \bank(k, A, N) * \bigast_{\substack{m=1 \\ m \neq i,j}}^{U} \bank(\mathcal{B}(expr_m, N), A, N)$$

Applying the disjointness axiom, the conflicting term $\bank(k, A, N) * \bank(k, A, N)$ immediately collapses to $\bot$. Because the separating conjunction distributes over logical false ($\bot * \Sigma \implies \bot$), the entire spatial state evaluates to a contradiction:
$$\cfu(E, A, N) \implies \bot$$

This unsatisfiable state explicitly and automatically catches the structural hazard. By reducing the detection of non-affine memory collisions to an SMT algebraic equivalence query that triggers a fundamental collapse in spatial logic, the framework definitively halts the synthesis of unsafe parallel Verilog without requiring strict polyhedral or affine geometric approximations.

\begin{figure}[htbp]
\centering
\begin{tikzpicture}[
    auto,
    >=stealth,
    block/.style = {
        rectangle, 
        draw=black, 
        thick,
        fill=blue!5, 
        text width=3.4cm, 
        align=center, 
        rounded corners=4pt, 
        minimum height=1.2cm,
        font=\sffamily\small
    },
    conflict/.style = {
        circle,
        draw=red!80!black,
        thick,
        fill=red!10,
        text width=1.8cm,
        align=center,
        font=\sffamily\scriptsize\bfseries
    },
    hw/.style = {
        rectangle,
        draw=black,
        thick,
        fill=gray!10,
        text width=2.5cm,
        align=center,
        minimum height=1.4cm,
        font=\sffamily\small
    },
    arrow/.style = {
        thick,
        ->
    }
]

    % ===================================================
    % PHASE 1: HARDWARE SCHEDULING CONFLICT
    % ===================================================
    \node[block, fill=green!5] (iter_i) at (0, 0) {
        \textbf{Iteration $i$ Access}\\
        \texttt{Offset expr}$_i$
    };
    
    \node[block, fill=green!5, below=0.8cm of iter_i] (iter_j) {
        \textbf{Iteration $j$ Access}\\
        \texttt{Offset expr}$_j$
    };

    \node[block, fill=orange!10, right=1.5cm of iter_i, yshift=-0.55cm, text width=2.8cm] (bank_k) {
        \textbf{Single-Port}\\
        \textbf{Memory Bank $k$}\\
        \scriptsize$\mathcal{B}(\text{expr}) = k$
    };

    % Adjusted position to 30% along the path and added vertical offsets to clear the lines completely
    \draw[arrow, red, thick] (iter_i.east) -- node[pos=0.5, above=0.15cm, font=\sffamily\scriptsize\bfseries, text=red] {Concurrent Access} (bank_k.west);
    \draw[arrow, red, thick] (iter_j.east) -- node[pos=0.3, below=0.15cm, font=\sffamily\scriptsize\bfseries, text=red] {} (bank_k.west);

    % ===================================================
    % PHASE 2: SPATIAL COLLAPSE
    % ===================================================
    \node[conflict, right=1.5cm of bank_k] (clash) {SPATIAL\\COLLAPSE};
    
    \node[block, fill=red!5, right=1.5cm of clash, text width=3.2cm] (logic) {
        \textbf{Logical Contradiction}\\
        \vspace{0.1cm}
        $bank(k) \ast bank(k)$\\
        $\Downarrow$\\
        \large $\perp$ \normalsize (False)
    };

    \draw[arrow, thick] (bank_k.east) -- (clash);
    \draw[arrow, thick] (clash) -- (logic);

    % ===================================================
    % PHASE 3: PERMISSION MAPPING & FALLBACK
    % ===================================================
    \node[block, fill=yellow!10, below=1.8cm of clash, text width=6.0cm] (policy) {
        \textbf{Permission Mapping $\mathcal{M}(\mathcal{S})$ Evaluation}\\
        \vspace{0.1cm}
        \footnotesize Status $\mathcal{S} \in \{\text{Unsat, Unknown, } \mathcal{T}_{out}\} \rightarrow$ \textbf{Sequential}
    };

    \draw[arrow, thick] (logic.south) |- (policy.east);

    % ===================================================
    % PHASE 4: RESULTING SERIALIZED HARDWARE RTL
    % ===================================================
    \node[hw, below=1.5cm of policy, xshift=-3.2cm] (cycle1) {
        \textbf{Clock Cycle 1}\\
        \vspace{0.05cm}
        \footnotesize Execute Iteration $i$\\
        \scriptsize (Bank $k$ Locked)
    };

    \node[hw, below=1.5cm of policy, xshift=3.2cm] (cycle2) {
        \textbf{Clock Cycle 2}\\
        \vspace{0.05cm}
        \footnotesize Execute Iteration $j$\\
        \scriptsize (Bank $k$ Available)
    };

    \node[draw=red, thick, fill=red!5, rounded corners=4pt, font=\sffamily\small\bfseries, inner sep=6pt, below=0.8cm of $(cycle1.south)!0.5!(cycle2.south)$] (mux) {
        MUX-Driven Stall Logic Installed
    };

    % Re-routed dashed lines into a clean, centralized organizational fork path
    \draw[arrow, thick, dashed] (policy.south) -- ++(0,-0.4) -| (cycle1.north);
    \draw[arrow, thick, dashed] (policy.south) -- ++(0,-0.4) -| (cycle2.north);
    
    \draw[arrow, red, thick] (cycle1.south) |- (mux.west);
    \draw[arrow, red, thick] (cycle2.south) |- (mux.east);

\end{tikzpicture}
\caption{Spatial Collapse and Fallback Logic: Concurrent bank collisions trigger an automatic logical contradiction ($\perp$), evaluating undecidable or conflicting layouts down to a safe, serialized sequential hardware structure.}
\label{fig:spatial_collapse_fallback}
\end{figure}
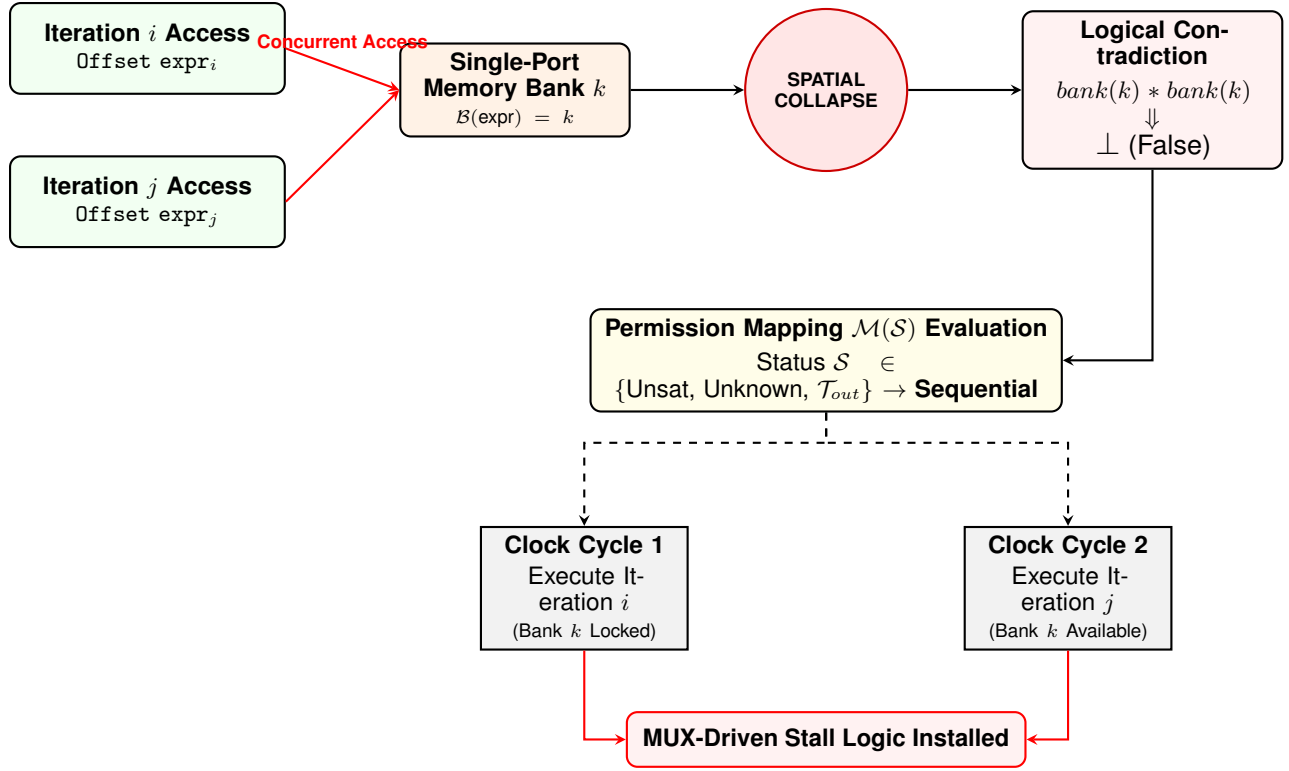

\section{Translating Spatial Disjointness to SMT}

\subsection{Algebraic Entailment Rules}

Standard Satisfiability Modulo Theories (SMT) solvers operate over decidable first-order theories (such as bitvectors and integer arithmetic). They do not natively understand the spatial heap $H$ or Separation Logic constructs like the separating conjunction ($\ast$). To bridge this gap, the spatial requirements of the hardware layout must be mechanically reduced into pure boolean algebra. We define a formal translation function, $\mathcal{T}$, which maps a spatial assertion into a pure first-order logic formula digestible by an SMT oracle, evaluated strictly over the Static Single Assignment (SSA) virtual registers accumulated in the path condition $\pi$.

The translation function acts upon the Conflict-Free Unrolling ($\cfu$) condition. When the execution of duplicated LLVM basic blocks attempts to union multiple bank predicates into $H$ (triggered by \texttt{load} and \texttt{store} instructions), the framework extracts the target bank indices. Given a path condition $\pi$, an array $A$ mapped in the static environment $\Gamma$, and a set of $U$ concurrent symbolic bank indices $K = \{k_1, k_2, \dots, k_U\}$, the spatial safety requirement $\pi \vdash \cfu(K, A, N)$ is translated as follows:

$$ \mathcal{T}(\pi \vdash \cfu(K, A, N)) \triangleq \pi \implies \bigwedge_{1 \le i < j \le U} (k_i \neq k_j) $$

This translation reduces the high-level spatial disjointness requirement to an equivalent matrix of pairwise inequality constraints over the symbolic SSA variables representing the bank indices. 

\paragraph{Mechanization of the Verification Condition (VC)}
The output of $\mathcal{T}$ yields the formal Verification Condition (VC) for the HLS scheduler. For example, under a cyclic partitioning directive, the bank index is derived from the flat 1D address offset computed by an LLVM \texttt{getelementptr} (GEP) instruction, such that $k = \text{offset} \bmod N$. A loop unrolled by a factor of $U = 3$ generating virtual registers for indices $\{k_1, k_2, k_3\}$ expands to the following algebraic constraint:

$$ \pi \implies \Big( (k_1 \neq k_2) \;\land\; (k_1 \neq k_3) \;\land\; (k_2 \neq k_3) \Big) $$

By embedding the path condition $\pi$ as the antecedent, the SMT solver evaluates these inequalities exclusively within the valid algebraic boundaries, branching constraints, and SSA assignments of the active execution trace. If the SMT solver proves that this VC is valid ($\text{Sat}$), it implies that no two SSA registers can map to the same physical bank identifier under the current path constraints. This algebraically preserves the core spatial invariant:

$$ \text{Valid}\left(\mathcal{T}(\pi \vdash \cfu(K, A, N))\right) \iff \langle \pi, H \rangle \not\models \bot $$

Consequently, because $\pi$ handles the pure LLVM IR algebra—including non-affine arithmetic such as modulo and division inherent to the bank mapping functions—the framework bypasses polyhedral geometry. Structural conflict detection is cleanly reduced to an SMT algebraic equivalence query over immutable SSA variables.

\subsection{Conservative Fallback for Undecidability}

A fundamental challenge in bypassing geometric polyhedral constraints via algebraic translation is that LLVM IR \texttt{getelementptr} (GEP) offset calculations and bank mapping functions (such as modulo or division) frequently generate non-linear integer arithmetic within the path condition $\pi$. Because first-order theories containing non-linear integer arithmetic are fundamentally undecidable, an automated SMT solver cannot be guaranteed to resolve the generated Verification Conditions within finite bounds. Consequently, the SMT oracle may return an \texttt{Unknown} status or trigger a deterministic synthesis timeout ($\mathcal{T}_{out}$).

While our spatial framework tries to resolve a subset of non-affine constraints that polyhedral models blindly serialize, non-linear integer arithmetic remains undecidable in first-order logic. To bridge this mathematical boundary and guarantee  structural safety in the physical hardware, our framework implements a deterministic sequential fallback mechanism. We mathematically formalize this operational boundary by augmenting the verification decision space. Let $\mathcal{S}$ be the verification status returned by the SMT oracle for a given translation $\mathcal{T}(\pi \vdash \cfu(K, A, N))$. The synthesized scheduler's permission mapping, $\mathcal{M}(\mathcal{S}) \in \{\text{Parallel}, \text{Sequential}\}$, is formally defined as:

$$
\mathcal{M}(\mathcal{S}) \triangleq 
\begin{cases} 
\text{Parallel} & \text{if } \mathcal{S} = \text{Sat} \\
\text{Sequential} & \text{if } \mathcal{S} = \text{Unsat} \lor \mathcal{S} = \text{Unknown} \lor \mathcal{S} = \mathcal{T}_{out}
\end{cases}
$$

By conservatively treating an \texttt{Unknown} or timed-out result identically to an \texttt{Unsat} result, the framework establishes a sound theoretical ceiling. It mechanically maps unresolvable non-linear algebra to a potential spatial hazard, explicitly triggering the Separation Logic contradiction ($P \ast P \implies \bot$).

\paragraph{Preserving the Soundness Invariant}
This mapping ensures that the framework does not under-approximates a structural hazard. From a hardware reality perspective, if the algebraic entailment engine cannot definitively prove the absolute mutual exclusion of the evaluated SSA bank indices ($k_i \neq k_j$), the framework assumes a worst-case scenario: that a physical memory conflict exists. 

The HLS scheduler reacts to $\text{Sequential}$ by immediately halting parallel synthesis for those duplicated basic blocks. It falls back to generating standard sequential execution states, inserting appropriate multiplexer-driven stall logic or pipeline registers to serialize the conflicting memory accesses over subsequent clock cycles. While this fallback occasionally yields a false positive (halting parallelization for an otherwise safe but undecidable non-linear GEP offset), it guarantees that the core Theorem of Soundness is not violated, ensuring the synthesized RTL is structurally immune to actual physical hardware collisions.

\section{Theorem of Soundness}

\subsection{Hardware Safety Guarantee}

To establish the structural safety of the synthesized hardware, we bridge the algebraic verification performed by the SMT oracle with the physical reality of the Register Transfer Level (RTL) execution trace. The core guarantee of our framework relies on the mathematical translation of spatial disjointness into pure first-order logic, proving that symbolic safety implies physical hardware safety.

\textbf{Theorem 1 (Hardware Safety via Spatial Soundness).} Let $E$ be a set of $U$ concurrent memory accesses targeting an array $A$ partitioned into $N$ banks, evaluated under a pure LLVM IR path condition $\pi$. Let $K = \{k_1, k_2, \dots, k_U\}$ be the set of symbolic bank indices extracted via the hardware mapping function $\mathcal{B}(expr_i, N)$. If the algebraic Verification Condition (VC) is validated by the SMT solver, then the spatial Separation Logic requirement strictly holds, guaranteeing that the synthesized RTL is free of structural memory bank collisions under all valid execution traces.

\subsection{Proof of Theorem 1}

\begin{proof}
Let $s : \text{Reg} \rightharpoonup \text{Val}$ be a valid monotonic store representing a specific execution trace, and let $s \models \pi$, meaning the store satisfies the pure algebraic path condition accumulated during the symbolic execution of the LLVM IR basic blocks.

Let $E$ be a set of $U$ concurrent memory operations, and let $K = \{k_1, k_2, \dots, k_U\}$ be the set of symbolic bank indices targeted by these operations, where each $k_i = \mathcal{B}(expr_i, N)$.

\textbf{Step 1: Algebraic Satisfaction.} 
By the premise of the theorem, the SMT solver has validated the Verification Condition (VC):
$$ \text{Valid}\left( \pi \implies \bigwedge_{1 \le i < j \le U} (k_i \neq k_j) \right) $$
Because $s \models \pi$, it mathematically follows by modus ponens that for all $1 \le i < j \le U$, the evaluations of the symbolic indices in the monotonic store yield strictly distinct values:
$$ \forall i, j \in [1, U], i \neq j \implies s(k_i) \neq s(k_j) $$

\textbf{Step 2: Spatial Expansion.}
The Conflict-Free Unrolling condition demands simultaneous ownership of all required banks in the spatial heap $H$:
$$ \langle \pi, H \rangle \models \cfu(K, A, N) \iff \langle \pi, H \rangle \models \bigast_{i=1}^{U} \bank(k_i, A, N) $$
By the semantics of the iterated separating conjunction, there must exist $U$ disjoint heaplets $h_1, h_2, \dots, h_U$ such that $H = \biguplus_{i=1}^U h_i$, where each heaplet satisfies $h_i \models \bank(k_i, A, N)$.

\textbf{Step 3: Disjointness and Absence of Collapse.}
Assume, for the sake of contradiction, that two concurrent memory operations $i$ and $j$ target the same physical memory bank, meaning $\text{dom}(h_i) \cap \text{dom}(h_j) \neq \emptyset$. 
By the definition of the static hardware environment $\Gamma(A) = (\alpha, M, \mathcal{P})$, the physical addresses bounded by the bank predicates are uniquely determined by the evaluated bank index $s(k)$. 
If the domains overlap, it strictly implies $s(k_i) = s(k_j)$. 
However, this directly contradicts the algebraic satisfaction proven in Step 1 ($s(k_i) \neq s(k_j)$). 
Thus, the spatial state avoids the fundamental Separation Logic contradiction ($P \ast P \implies \bot$); the spatial heap $H$ is satisfiable and strictly disjoint:
$$ \forall i \neq j, \quad h_i \perp h_j \iff \text{dom}(h_i) \cap \text{dom}(h_j) = \emptyset $$

\textbf{Step 4: Hardware Synthesis Translation.}
Because the spatial domains of all $U$ bank predicates are mutually disjoint within the physical address space $\mathbb{L}$, the concurrent LLVM \texttt{load} or \texttt{store} instructions assert ownership over entirely independent physical SRAM interfaces. 
Consequently, when the HLS scheduler lowers these LLVM IR instructions into parallel Verilog control signals, no two signals will attempt to assert the enable (EN) and write-enable (WE) pins of the same single-ported physical memory bank $k$ during the same hardware clock cycle. 

Therefore, validating the algebraic Verification Condition provides a mathematically sound proof that the resulting synthesized RTL execution trace is free of structural memory collisions.
\end{proof}

\section{Conclusion}

\subsection{Summary of Contributions}
High-Level Synthesis (HLS) relies on loop unrolling and array partitioning to maximize parallel hardware throughput. However, traditional dependency solvers based on the geometric polyhedral model over-approximate when scheduling concurrent memory accesses governed by non-affine array indices. 
To overcome this bottleneck, this paper presented a theoretical framework that repurposes Separation Logic (SL) to statically model physical hardware memory banks as distinct spatial resources directly over the LLVM Intermediate Representation (IR). 
By defining the Conflict-Free Unrolling (CFU) condition utilizing the iterated separating conjunction, we reduced the detection of structural memory hazards to an algebraic, SMT-solvable constraint matrix. 
This approach translates spatial disjointness into pairwise inequalities across immutable Static Single Assignment (SSA) registers, 
bypassing the affine limitations of HLS compilers and guaranteeing the structural safety of the synthesized RTL.

\subsection{Future Work}
Looking forward, we aim to extend this theoretical foundation in both formal and practical directions. 
Theoretically, we plan to rigorously define Separation Logic loop invariants to effectively manage the spatial state-space of partially unrolled loops and  pipelined memory architectures, expanding the framework's applicability to a broader class of complex scheduling scenarios. 
Furthermore, we intend to refine the spatial heap model by incorporating fractional permissions in Concurrent Separation Logic \cite{bornat2005permission,brookes2016concurrent} to  differentiate between concurrent LLVM \texttt{load} and \texttt{store} instructions. 
While the current framework enforces strict exclusive ownership for all memory operations to safely model single-ported SRAM constraints, adopting fractional permissions will allow shared spatial ownership for read-only accesses. 
This theoretical extension will prevent artificial spatial collapses during safe concurrent reads, optimizing throughput for multi-ported memory or ROM broadcast architectures.

Practically, the  next step is the software realization of this  framework as a custom LLVM analysis pass, integrated natively with an automated SMT oracle such as Z3 \cite{de2008z3}. 
This implementation will allow us to empirically benchmark the framework's conflict detection accuracy, compilation time overhead, and resulting hardware resource efficiency against standard industry HLS tools, bridging the gap between formal programming language theory and physical hardware generation.

%Bibliography
\bibliographystyle{unsrt}  
\bibliography{references}

@article{hoare1969axiomatic,
  title={An axiomatic basis for computer programming},
  author={Hoare, Charles Antony Richard},
  journal={Communications of the ACM},
  volume={12},
  number={10},
  pages={576--580},
  year={1969},
  publisher={ACM New York, NY, USA}
}

@inproceedings{reynolds2002separation,
  title={Separation logic: A logic for shared mutable data structures},
  author={Reynolds, John Charles},
  booktitle={Proceedings 17th Annual IEEE Symposium on Logic in Computer Science},
  pages={55--74},
  year={2002},
  organization={IEEE}
}

@article{brookes2016concurrent,
  title={Concurrent separation logic},
  author={Brookes, Stephen and O'Hearn, Peter W},
  journal={ACM SIGLOG News},
  volume={3},
  number={3},
  pages={47--65},
  year={2016},
  publisher={ACM New York, NY, USA}
}

@inproceedings{o2004resources,
  title={Resources, concurrency and local reasoning},
  author={O’hearn, Peter W},
  booktitle={International Conference on Concurrency Theory},
  pages={49--67},
  year={2004},
  organization={Springer}
}

@inproceedings{berdine2005symbolic,
  title={Symbolic execution with separation logic},
  author={Berdine, Josh and Calcagno, Cristiano and O’hearn, Peter W},
  booktitle={Asian Symposium on Programming Languages and Systems},
  pages={52--68},
  year={2005},
  organization={Springer}
}

@article{winterstein2015separation,
  title={Separation logic for high-level synthesis},
  author={Winterstein, Felix J and Bayliss, Samuel R and Constantinides, George A},
  journal={ACM Transactions on Reconfigurable Technology and Systems (TRETS)},
  volume={9},
  number={2},
  pages={1--23},
  year={2015},
  publisher={ACM New York, NY, USA}
}

@article{he2021type,
  title={A type system for extracting functional specifications from memory-safe imperative programs},
  author={He, Paul and Westbrook, Eddy and Carmer, Brent and Phifer, Chris and Robert, Valentin and Smeltzer, Karl and {\c{S}}tef{\u{a}}nescu, Andrei and Tomb, Aaron and Wick, Adam and Yacavone, Matthew and others},
  journal={Proceedings of the ACM on Programming Languages},
  volume={5},
  number={OOPSLA},
  pages={1--29},
  year={2021},
  publisher={ACM New York, NY, USA}
}

@article{villard2013here,
  title={{Here be wyverns! Verifying LLVM bitcode with llStar}},
  author={Villard, Jules},
  journal={Unpublished manuscript at \url{https://jvillard.net/pub/llstar-draft-oct13.pdf}},  
  year={2013}
}

@article{cong2011high,
  title={High-level synthesis for FPGAs: From prototyping to deployment},
  author={Cong, Jason and Liu, Bin and Neuendorffer, Stephen and Noguera, Juanjo and Vissers, Kees and Zhang, Zhiru},
  journal={IEEE Transactions on Computer-Aided Design of Integrated Circuits and Systems},
  volume={30},
  number={4},
  pages={473--491},
  year={2011},
  publisher={IEEE}
}

@inproceedings{bornat2005permission,
  title={Permission accounting in separation logic},
  author={Bornat, Richard and Calcagno, Cristiano and O'Hearn, Peter and Parkinson, Matthew},
  booktitle={Proceedings of the 32nd ACM SIGPLAN-SIGACT symposium on Principles of programming languages},
  pages={259--270},
  year={2005}
}

@inproceedings{de2008z3,
  title={Z3: An efficient SMT solver},
  author={De Moura, Leonardo and Bj{\o}rner, Nikolaj},
  booktitle={International conference on Tools and Algorithms for the Construction and Analysis of Systems},
  pages={337--340},
  year={2008},
  organization={Springer}
}

@article{nane2015survey,
  title={A survey and evaluation of FPGA high-level synthesis tools},
  author={Nane, Razvan and Sima, Vlad-Mihai and Pilato, Christian and Choi, Jongsok and Fort, Blair and Canis, Andrew and Chen, Yu Ting and Hsiao, Hsuan and Brown, Stephen and Ferrandi, Fabrizio and others},
  journal={IEEE Transactions on Computer-Aided Design of Integrated Circuits and Systems},
  volume={35},
  number={10},
  pages={1591--1604},
  year={2015},
  publisher={IEEE}
}

@article{canis2013legup,
  title={LegUp: An open-source high-level synthesis tool for FPGA-based processor/accelerator systems},
  author={Canis, Andrew and Choi, Jongsok and Aldham, Mark and Zhang, Victor and Kammoona, Ahmed and Czajkowski, Tomasz and Brown, Stephen D and Anderson, Jason H},
  journal={ACM Transactions on Embedded Computing Systems (TECS)},
  volume={13},
  number={2},
  pages={1--27},
  year={2013},
  publisher={ACM New York, NY, USA}
}

@inproceedings{josipovic2018dynamically,
  title={Dynamically scheduled high-level synthesis},
  author={Josipovi{\'c}, Lana and Ghosal, Radhika and Ienne, Paolo},
  booktitle={Proceedings of the 2018 ACM/SIGDA International Symposium on Field-Programmable Gate Arrays},
  pages={127--136},
  year={2018}
}

@inproceedings{wang2013memory,
  title={Memory partitioning for multidimensional arrays in high-level synthesis},
  author={Wang, Yuxin and Li, Peng and Zhang, Peng and Zhang, Chen and Cong, Jason},
  booktitle={Proceedings of the 50th Annual Design Automation Conference},
  pages={1--8},
  year={2013}
}

@article{liu2017polyhedral,
  title={Polyhedral-based dynamic loop pipelining for high-level synthesis},
  author={Liu, Junyi and Wickerson, John and Bayliss, Samuel and Constantinides, George A},
  journal={IEEE Transactions on Computer-Aided Design of Integrated Circuits and Systems},
  volume={37},
  number={9},
  pages={1802--1815},
  year={2017},
  publisher={IEEE}
}

@inproceedings{bondhugula2008practical,
  title={A practical automatic polyhedral parallelizer and locality optimizer},
  author={Bondhugula, Uday and Hartono, Albert and Ramanujam, Jagannathan and Sadayappan, Ponnuswamy},
  booktitle={Proceedings of the 29th ACM SIGPLAN Conference on Programming Language Design and Implementation},
  pages={101--113},
  year={2008}
}

\end{document}